\documentclass{article}
\usepackage{geometry}
\usepackage{amsmath}
\usepackage{graphicx}
\usepackage{esint}

\makeatletter
\newcommand{\lyxaddress}[1]{
\par {\raggedright #1
\vspace{1.4em}
\noindent\par}
}

\usepackage{wasysym}
\usepackage{cite}

\usepackage{chemarr}

\makeatother

\begin{document}

\title{Optimal Temporal Patterns for Dynamical Cellular Signaling}

\author{Yoshihiko Hasegawa}

\date{\today}

\maketitle

\lyxaddress{Department of Information and Communication Engineering, Graduate
School of Information Science and Technology, The University of Tokyo,
Tokyo 113-8656, Japan}
\begin{abstract}
Cells use temporal dynamical patterns to transmit information via
signaling pathways. As optimality with respect to the environment
plays a fundamental role in biological systems, organisms have evolved
optimal ways to transmit information. Here, we use optimal control
theory to obtain the dynamical signal patterns for the optimal transmission
of information, in terms of efficiency (low energy) and reliability
(low uncertainty). Adopting an activation-deactivation decoding network,
we reproduce several dynamical patterns found in actual signals, such
as steep, gradual, and overshooting dynamics. Notably, when minimizing
the energy of the input signal, the optimal signals exhibit overshooting,
which is a biphasic pattern with transient and steady phases; this
pattern is prevalent in actual dynamical patterns. We also identify
conditions in which these three patterns (steep, gradual, and overshooting)
confer advantages. Our study shows that cellular signal transduction
is governed by the principle of minimizing free energy dissipation
and uncertainty; these constraints serve as selective pressures when
designing dynamical signaling patterns. 
\end{abstract}

\section{Introduction}

\begin{figure*}
\begin{centering}
\includegraphics[width=16cm]{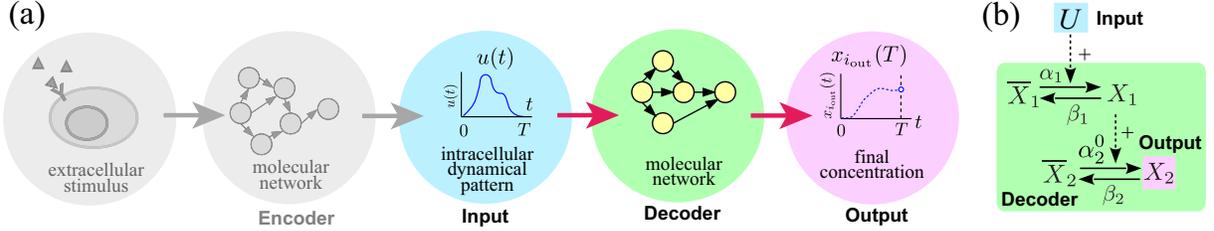}
\par\end{centering}

\protect\caption{Overview of dynamical signal transduction. (a) An extracellular stimulus
is encoded into an intracellular dynamical pattern. The decoder reads
the dynamical pattern and outputs the result as the final concentration.
This study focuses on the latter process (colored flow). The intracellular
dynamical pattern $u(t)$ and the final concentration $x_{i_{\mathrm{out}}}(T)$
correspond to input and output, respectively, and we optimize the
dynamical pattern. (b) Two-molecule decoding network (the two-stage
model), which reports the result as $X_{2}$. $\overline{X}_{i}$
and $X_{i}$ ($i=1,2$) denote the inactive and active molecules,
respectively. \label{fig:model} }
\end{figure*}

Cells transmit information through signal transduction and transcription
networks \cite{Alon:2007:NetMotif,Alon:2007:SystBiolBook}. Recent
studies have revealed that, along with the identity and static concentration
of molecules, cells also encode information into dynamical patterns
\cite{Behar:2010:TemporalCodes,Kubota:2012:InsulinAKT,Purvis:2012:p53,Purvis:2013:SignalingReview,Sonnen:2014:SignalingReview,Selimkhanov:2014:DynSig,Lin:2015:CombGRN}.
Examples of dynamical patterns include extracellular signal-regulated
kinase (ERK), the yeast transcription factor Msn2, the transcription
factor NF-$\kappa$B, a protein kinase AKT, and calcium signaling.
Many studies have used nonlinear and stochastic approaches to investigate
the properties of dynamical cellular information processing \cite{Tostevin:2009:MI,Mora:2010:MLE,Mugler:2010:OscSignal,Hansen:2013:DynDec,Kobayashi:2010:PRL,McMahon:2015:MI,Becker2015:CellPrediction,Makadia:2015:DynamicSignal}.
Because signal transduction plays central and crucial roles in the
survival of cells, the time course of dynamical patterns is expected
to be highly optimized so that cells can efficiently and accurately
transmit information. Although the advantages of dynamical signals
over static ones have been extensively studied \cite{Tostevin:2012:AFM,Selimkhanov:2014:DynSig},
there has been little investigation into determining which dynamical
signals are the best. We assume that two principles that are prevalent
in many biological systems govern the optimality of signal patterns:
energetic efficiency (low energy) and reliability (low uncertainty).
Biological systems are often characterized by low energy consumption.
For instance, neuronal systems are known to function with remarkably
low energy consumption. Specifically, in neurons, information processing
capability is bounded by the amount of energy consumption and it is
reported that the energy consumption of the brain has limited its
size (\cite{Laughlin:2001:Energy,Sengupta:2014:PC} and references
therein). Biochemical networks process information for a variety of
purposes, and higher specificity, lower variation, and larger signal
amplification demand more energy consumption \cite{Mehta:2016:InfoBioReview}.
These facts induce us to think that the energetic cost also plays
important roles in information transmission of the dynamical signal
transduction. A major cause of interference with reliability is molecular
noise, which degrades the quality of transmitted information. Despite
the stochastic nature of cellular processes, organisms have acquired
several mechanisms to resist or to take advantage of noise in order
to enhance biological functionalities \cite{McDonnell:2008:SRBook}.
As these two principles are of significance, the dynamical transmission
of information has evolved in such a way that it optimally satisfies
these principles. We divide the dynamical signal transduction into
two parts: encoding of extracellular stimuli into intracellular dynamical
patterns, and decoding of the dynamical patterns into the response
(in the present manuscript, the response corresponds to the concentration
of output molecular species) (Fig.~\ref{fig:model}(a)). By viewing
the dynamical signal as an input, the decoding network as the system
to be controlled, and the output concentration as an output (Fig.~\ref{fig:model}(a)),
we use optimal control theory \cite{Kamien:2012:DynOpt,Hull:2013:Optimal}
to determine the signal dynamics that optimize energy efficiency and
the reliability of transmitted information. We quantify the energetic
cost by free energy dissipation when generating temporal dynamical
patterns and the uncertainty by variance of output molecular species
concentration. To decode the dynamical signals, we adopt an activation-deactivation
network (Fig.~\ref{fig:model}(b)), which is a motif commonly used
in biochemical networks. We optimize the dynamical patterns, i.e.
the input. Decoders (decoding networks) may also coevolve to maximally
reading out information from signals, but in this paper, we focus
on optimizing the input. No matter how precisely the decoder is able
to read dynamical signals, it is impossible to totally eliminate the
uncertainty due to the inherent stochasticity. Therefore, there is
a lower bound on the uncertainty and the bound is determined by input. 

From our calculations, we identify three basic patterns for the signals:
steep (Fig.~\ref{fig:Phos_U}(a)), gradual (Fig.~\ref{fig:Phos_U}(b)),
and overshooting (Fig.~\ref{fig:Phos_U}(c)). We show that the steep
pattern minimizes the energy, whereas the gradual pattern minimizes
the uncertainty. Intriguingly, when minimizing the energy of a dynamical
pattern while achieving a higher output concentration, overshooting
is the optimal pattern; this pattern can often be seen in real processes.
We identify the conditions in which these three patterns (steep, gradual,
and overshooting) confer advantages. We note that these patterns are
prevalent in signal transduction, and our calculations show that minimizing
the energetic cost and uncertainty plays important roles in the evolutionary
design of dynamical signaling patterns.

\section{Methods}

\subsection{Model}

Dynamical signal transduction is typically separated into two parts
\cite{Sonnen:2014:SignalingReview}: encoding of extracellular stimuli
into intracellular dynamical patterns, and decoding of the dynamical
patterns into the response (Fig.~\ref{fig:model}(a)); our study
focuses on the latter process, and dynamical signals are optimized
for readout by a particular decoder, where the optimization criteria
are to minimize the energy consumed by signal generation and to minimize
the uncertainty of the readout. 

In cells, dynamical signals are decoded by molecular networks. We
consider a molecular network consisting of $N$ molecular species
$X_{1},X_{2},...,X_{N}$ and we define $x_{i}$ as the concentration
of $X_{i}$. The input signal is carried by a molecular species $U$,
whose concentration $u(t)$ follows a dynamic pattern, and its onset
is $t=0$. In our analysis, we use optimal control theory to optimize
the temporal pattern of $u(t)$. Let the $i_{\mathrm{out}}$th molecular
species $X_{i_{\mathrm{out}}}$ be the output of the network. The
network reads the information from the input $u(t)$ (i.e., intracellular
dynamical signal) and outputs the result as the concentration of $X_{i_{\mathrm{out}}}$
at time $T$ ($T>0$) (Fig.~\ref{fig:model}(a)), i.e., $x_{i_{\mathrm{out}}}(T)$
carries information about the input signal. 

Consider the evolutionary design of the dynamical signal $u(t)$ that
attains the desired concentration of an output molecule $X_{i_{\mathrm{out}}}$
at $t=T$. Although there might be many possible dynamics for $u(t)$
($0\le t\le T$) that result in the desired output concentration,
the most biologically preferable ones are selected. We can expect
that the signals with lower energetic cost will be selected. In addition,
biochemical reactions are subject to noise, due to the smallness of
the cells. The noise degrades the information, and hence transmission
with lower uncertainty is desirable. Considering the energy of the
input and the uncertainty of the concentration of the output molecule,
we wish to find a signal $u(t)$ that minimizes a performance index
$R$, defined as 
\begin{equation}
R=\gamma_{i_{\mathrm{out}}}(T)+w\Pi,\label{eq:J_def1}
\end{equation}
where $\gamma_{i}(t)=\left\langle \left(x_{i}(t)-\mu_{i}(t)\right)^{2}\right\rangle $
is the variance of the concentration of the $i$th molecule at time
$t$ {[}$\mu_{i}(t)=\left\langle x_{i}(t)\right\rangle $ is the mean{]}
which quantifies the uncertainty, $\Pi$ is the energetic cost of
the signal $u(t)$, and $w$ is a weight parameter in the range $0<w<\infty$,
which represents the importance of the energy for the performance
index. As denoted, the output $X_{i_{\mathrm{out}}}$ has the target
concentration at time $t=T$. Therefore, the mean concentration of
the output $X_{i_{\mathrm{out}}}$, which we denote as $\mu_{i_{\mathrm{out}}}(t)$,
must attain the predefined target concentration $\mu_{i_{\mathrm{out}}}^{\mathrm{trg}}$
(``trg'' is short for ``target'') at time $t=T$, i.e., 
\begin{equation}
\mu_{i_{\mathrm{out}}}(T)=\mu_{i_{\mathrm{out}}}^{\mathrm{trg}}\label{eq:endpoint_constraint}
\end{equation}
is a boundary condition.

\subsection{Quantification of uncertainty}

We quantify the uncertainty as the variance of the output molecular
species concentration $\gamma_{i_{\mathrm{out}}}(T)$; the derivation
is shown below. The dynamics of molecular networks, which decode $u(t)$
and output the result, can be generally captured by the following
rate equation: 
\[
\dot{x}_{i}(t)=\sum_{\ell=1}^{N_{r}}s_{i\ell}v_{\ell}(\boldsymbol{x},u),
\]
where $\boldsymbol{x}=(x_{1},x_{2},...,x_{N})$, $\{s_{i\ell}\}=\boldsymbol{S}$
is a stoichiometry matrix, $v_{\ell}(\boldsymbol{x},u)$ is the reaction
velocity of the $\ell$th reaction, and $N_{r}$ is the number of
reactions. Due to the smallness of the cells, chemical reactions are
subject to stochasticity. We describe the noisy dynamics by the  Fokker--Planck
equation (FPE) \cite{Gillespie:2000:CLE,Klipp:2013:SystemsBiology}:
\begin{equation}
\frac{\partial}{\partial t}P(\boldsymbol{x};t)=-\sum_{i}\frac{\partial}{\partial x_{i}}\sum_{\ell}s_{i\ell}v_{\ell}(\boldsymbol{x},u)P(\boldsymbol{x};t)+Q\sum_{i,j}\frac{\partial^{2}}{\partial x_{i}\partial x_{j}}\sum_{\ell}s_{i\ell}s_{j\ell}v_{\ell}(\boldsymbol{x},u)P(\boldsymbol{x};t),\label{eq:FPE_def}
\end{equation}
where $P(\boldsymbol{x};t)$ is the probability density of $\boldsymbol{x}$
at time $t$, and $Q$ is the noise intensity related to the volume
$V$ via $Q=(2V)^{-1}$. Optimal control theory and related variational
methods have been employed by many researchers \cite{Forger:2004:OptimalControl,Moehlis:2006:Optimal,Hasegawa:2013:OptimalPRC,Hasegawa:2014:PRL,Iolov:2014:OptimalControl}.
Although stochastic optimal control theory has been applied in biological
contexts \cite{Iolov:2014:OptimalControl}, it is difficult to apply
it to multivariate models. Instead, we describe the dynamics by using
the time evolution of moments derived from Eq.~\eqref{eq:FPE_def}
\cite{Rodriguez:1996:SpikeNeurons,Tuckwell:2009:HHeq}. For general
nonlinear models, the naive calculation of moment equations results
in an infinite hierarchy of differential equations. Because our adopted
models are linear with respect to $\boldsymbol{x}$ (cf. Eqs.~\eqref{eq:ODE_1}
and \eqref{eq:ODE_2}), we can obtain closed differential equations
(see the supplementary material). For moments of up to the second
order {[}mean $\mu_{i}(t)$, variance $\gamma_{i}(t)$, and covariance
$\rho_{ij}=\left\langle \left(x_{i}-\mu_{i}\right)\left(x_{j}-\mu_{j}\right)\right\rangle ${]},
we have the following moment equation with respect to $\boldsymbol{z}=(\mu_{1},...,\mu_{n},\gamma_{1},...,\gamma_{n},\rho_{12},\rho_{23},...,\rho_{1n})$:
\begin{equation}
\dot{z}_{i}(t)=h_{i}(\boldsymbol{z},u),\label{eq:ODE_def}
\end{equation}
where $h_{i}(\boldsymbol{z},u)$ is right-hand side of the moment
equation and the dimensionality of $\boldsymbol{z}$ is $M=N(N+3)/2$.
 With the moment equation \eqref{eq:ODE_def}, we can reduce the
stochastic optimal control problem to a deterministic one. 

\begin{figure}
\begin{centering}
\includegraphics[width=15cm]{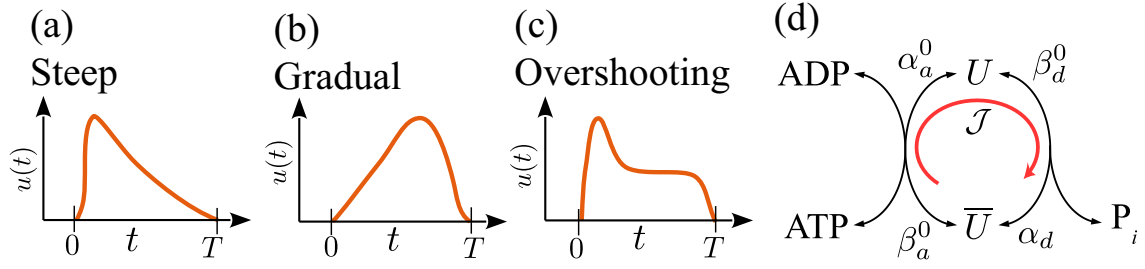}
\par\end{centering}

\protect\caption{(a)--(c) Dynamical signal $u(t)$ ($0\le t\le T$) as a function of
$t$ for the three typical patterns: (a) steep, (b) gradual, and (c)
overshooting. (d) Activation-deactivation of signaling molecule $U$.
Inactive molecule $\overline{U}$ is activated to become active molecule
$U$, which is mediated by phosphorylation, and deactivation is mediated
by dephosphorylation. Net flux $\mathcal{J}$ occurs in the clockwise
direction (shown by an arrow). \label{fig:Phos_U}}
\end{figure}

\subsection{Quantification of energetic cost}

\begin{figure}
\begin{centering}
\includegraphics[width=12cm]{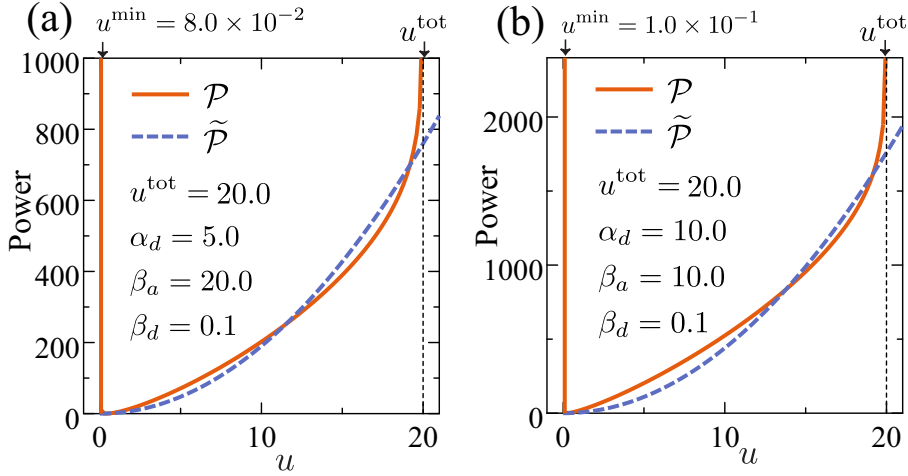} 
\par\end{centering}

\protect\caption{Exact representation of the power $\mathcal{P}$ (solid line) and
its quadratic approximation $\widetilde{\mathcal{P}}$ (dashed line).
(a) $u^{\mathrm{tot}}=20.0$, $\alpha_{d}=5.0$, $\beta_{a}=20.0$,
and $\beta_{d}=0.1$ ($q=1.9$ for quadratic); and (b) $u^{\mathrm{tot}}=20.0$,
$\alpha_{d}=10.0$, $\beta_{a}=10.0$, and $\beta_{d}=0.1$ ($q=4.4$
for quadratic). We use $k_{B}=1$ and $\mathcal{T}=1$, without loss
of generality. \label{fig:exact_vs_quadratic}}
\end{figure}

We next define the energetic cost of a signal ($\Pi$ in Eq.~\eqref{eq:J_def1})
as the free energy dissipated by controlling the concentration $u(t)$
such that it follows the desired temporal dynamics. We derive the
energetic cost of the signal $U$ with a simple biochemical model
after Ref.~\cite{Qian:2005:Thermodynamics,Qian:2007:PiReview,Beard:2008:ChemicalBook}.
We assume that input molecular species $U$ is activated from $\overline{U}$
and undergoes the following reaction: 
\begin{equation}
\overline{U}\rightleftharpoons U.\label{eq:reaction_def}
\end{equation}
Note that the total concentration $u^{\mathrm{tot}}=u+\overline{u}$
does not change with time, where $u$ and $\overline{u}$ are the
concentrations of $U$ and $\overline{U}$, respectively. In general,
the deactivation reaction is not the reverse of activation. Therefore,
when considering the energetic cost of generating $U$, we have to
handle activation and deactivation separately. Activation is typically
mediated by phosphorylation, where $\overline{U}$ is activated by
a kinase through the transfer of phosphate from ATP. The deactivation
is mediated by dephosphorylation, in which a phosphatase transfers
inorganic phosphate ($\mathrm{P}_{i}$) to the solution. These reactions
are written as
\begin{equation}
\overline{U}+\mathrm{ATP}\xrightleftharpoons[\beta_{a}^{0}]{\alpha_{a}^{0}}U+\mathrm{ADP},\hspace*{1em}U\xrightleftharpoons[\beta_{d}^{0}]{\alpha_{d}}\overline{U}+\mathrm{P}_{i},\label{eq:U_reaction}
\end{equation}
where $\alpha_{a}^{0}$, $\beta_{a}^{0}$, $\alpha_{d}$, and $\beta_{d}^{0}$
are reaction rates. When these parameters are held constant and the
system is closed, the system relaxes to an equilibrium state. Let
$c_{T}$, $c_{D}$, and $c_{P}$ be the concentrations of ATP, ADP,
and $\mathrm{P}_{i}$, respectively. In the natural cellular environment
where the system is open, $c_{T}$, $c_{D}$, and $c_{P}$ can be
regarded as constant, due to external agents \cite{Qian:2007:PiReview},
which we denote as $c_{T}^{0}$, $c_{D}^{0}$, and $c_{P}^{0}$, respectively.
Therefore the system relaxes to a nonequilibrium steady state (NESS).
The steady-state concentration of $U$ is
\begin{equation}
u^{\mathrm{ss}}=\frac{u^{\mathrm{tot}}\left(\beta_{d}+\alpha_{a}\right)}{\beta_{a}+\beta_{d}+\alpha_{a}+\alpha_{d}},\label{eq:uss_def}
\end{equation}
where $\beta_{a}=\beta_{a}^{0}c_{D}^{0}$, $\alpha_{a}=\alpha_{a}^{0}c_{T}^{0}$,
and $\beta_{d}=\beta_{d}^{0}c_{P}^{0}$. At the steady state, the
net flux (clockwise direction in Fig.~\ref{fig:Phos_U}(d)) is $\mathcal{J}^{\mathrm{ss}}=u^{\mathrm{tot}}\left(\alpha_{a}\alpha_{d}-\beta_{a}\beta_{d}\right)/\left(\beta_{a}+\beta_{d}+\alpha_{a}+\alpha_{d}\right)$.
The free energy dissipated during one cycle (i.e., $\overline{U}\rightarrow U\rightarrow\overline{U}$
in the clockwise direction in Fig.~\ref{fig:Phos_U}(d)) is $\Delta\phi=k_{B}\mathcal{T}\ln\left\{ \beta_{a}\beta_{d}/\left(\alpha_{a}\alpha_{d}\right)\right\} $,
where $k_{B}$ is the Boltzmann constant and $\mathcal{T}$ is the
temperature (see the supplementary material). Therefore the instantaneous
free-energy dissipation (i.e., power) is 
\begin{equation}
\mathcal{P}=-\mathcal{J}^{\mathrm{ss}}\Delta\phi=\frac{k_{B}\mathcal{T}u^{\mathrm{tot}}\left(\alpha_{a}\alpha_{d}-\beta_{a}\beta_{d}\right)}{\beta_{a}+\beta_{d}+\alpha_{a}+\alpha_{d}}\ln\left(\frac{\alpha_{a}\alpha_{d}}{\beta_{a}\beta_{d}}\right).\label{eq:P_def_exact}
\end{equation}
Equation~\eqref{eq:P_def_exact} quantifies the cost of the activation-deactivation
of $U$. Next, in order to yield the dynamics of $U$, we assume
that kinase activity is controlled by an upstream molecular species
and thus $\alpha_{a}$ varies temporally. We assume that the relaxation
of Eq.~\eqref{eq:U_reaction} is very fast {[}i.e., $1/(\beta_{a}+\beta_{d}+\alpha_{a}+\alpha_{d})$
is very small{]} so that the concentration $u$ is well approximated
by $u^{\mathrm{ss}}$ (Eq.~\eqref{eq:uss_def}) even for the case
of a time-varying $\alpha_{a}$. From Eq.~\eqref{eq:uss_def}, $\alpha_{a}$
can be represented as a function of $u$:
\begin{equation}
\alpha_{a}=\frac{u}{u^{\mathrm{tot}}-u}\left(\alpha_{d}+\beta_{a}\right)-\beta_{d}.\label{eq:alpha_a_as_u}
\end{equation}
From the condition $\alpha_{a}>0$, the minimum of $u$ is $u^{\min}=\beta_{d}u^{\mathrm{tot}}/\left(\beta_{a}+\beta_{d}+\alpha_{d}\right)$.
For this dynamic case, the instantaneous free energy dissipation
is given as a function of $u$ which is Eq.~\eqref{eq:P_def_exact}
along with Eq.~\eqref{eq:alpha_a_as_u}. The molecular species upstream
from $U$ also consumes energy; however, signal transduction generally
amplifies the external stimuli, and so the concentration of the upstream
molecular species is less than the concentration of $U$ \cite{Aksan:2004:Kinetic,Thomson:2011:SignalNum}.
Furthermore, we assume that the concentration of the molecular species
downstream from $U$ is also less than that of $U$ \cite{Aksan:2004:Kinetic,Thomson:2011:SignalNum}.
Indeed, in the ERK pathway, the concentration of ERK is higher than
that of its downstream molecular species (that is, the energetic cost
of decoding $U$ is smaller than the cost of generating it). Therefore,
we assume that the energetic cost of generating $U$ (i.e., Eq.~\eqref{eq:P_def_exact})
dominates the overall energetic cost. The free energy dissipated during
$0\le t\le T$ is 
\begin{equation}
\Pi=\int_{0}^{T}\mathcal{P}dt.\label{eq:Pi_exact}
\end{equation}
Because it is difficult to use the exact representation of $\Pi$
in the optimal control calculation, we approximate $\mathcal{P}$
with a simpler expression: $\mathcal{P}=0$ at a zero-flux point $u=u^{\mathrm{zf}}=\beta_{d}u^{\mathrm{tot}}/(\beta_{d}+\alpha_{d})$
where $\mathcal{J}$ vanishes. Assuming that $u^{\mathrm{zf}}$ is
sufficiently small, $\mathcal{P}=0$ when $u$ is very low concentration.
Furthermore, $\mathcal{P}$ increases superlinearly as $u$ increases
from the zero-flux point $u=u^{\mathrm{zf}}$. Taking into account
the conditions and computational feasibility, we use the approximation
$\mathcal{P}\simeq\widetilde{\mathcal{P}}$ with $\mathcal{\widetilde{\mathcal{P}}}=qu^{2}$,
where $q$ is a proportionality coefficient. Then, the free energy
dissipation during the period $0\le t\le T$ is approximated by 
\begin{equation}
\widetilde{\Pi}=\int_{0}^{T}\widetilde{\mathcal{P}}dt=\int_{0}^{T}qu(t)^{2}dt.\label{eq:Pi_quadratic}
\end{equation}
Figure~\ref{fig:exact_vs_quadratic} compares the exact expression
of $\mathcal{P}$ (Eq.~\eqref{eq:P_def_exact} along with Eq.~\eqref{eq:alpha_a_as_u})
with its quadratic approximation $\widetilde{\mathcal{P}}$ for two
settings; the exact and quadratic results are shown by solid and dashed
lines, respectively. For both parameter settings, we see that the
behavior of the quadratic approximation is similar to that of the
exact one. The major difference between the exact and the quadratic
representations is that Eq.~\eqref{eq:P_def_exact} diverges to $\infty$
for $u\rightarrow u^{\min}$ and $u\rightarrow u^{\mathrm{tot}}$.
Therefore, in order for the quadratic expression to well approximate
the exact energetic cost, $u^{\mathrm{tot}}$ should satisfy requirements
in addition to the condition of $u^{\mathrm{zf}}$. If $u^{\mathrm{tot}}$
is too small, the energy divergence at $u=u^{\mathrm{tot}}$ prevents
the signal to have higher peaks. On the other hand, if $u^{\mathrm{tot}}$
is excessively large, the exact energetic cost becomes almost linear
with respect to $u$ which makes the quadratic approximation less
reliable. With this approximation, we set $u(t)=0$ for $t<0$ and
$u(t)\ge0$ for $t>0$ (we define that the onset is the time when
$u(t)$ becomes positive). 

\subsection{Finding the optimal signaling pattern}

We wish to obtain the optimal control $u(t)$ that minimizes $R$
of Eq.~\eqref{eq:J_def1} while satisfying Eq.~\eqref{eq:ODE_def}
and the predefined target mean concentration of Eq.~\eqref{eq:endpoint_constraint}
{[}$\mu_{i_{\mathrm{out}}}(T)=z_{i_{\mathrm{out}}}(T)=\mu_{i_{\mathrm{out}}}^{\mathrm{trg}}${]}.
Then, by virtue of optimal control theory \cite{Kamien:2012:DynOpt,Hull:2013:Optimal},
we minimize the following augmented performance index:
\begin{align}
\widetilde{R} & =\gamma_{i_{\mathrm{out}}}(T)+w\widetilde{\Pi}+\nu\left(z_{i_{\mathrm{out}}}(T)-\mu_{i_{\mathrm{out}}}^{\mathrm{trg}}\right)+\sum_{i=1}^{M}\int_{0}^{T}\lambda_{i}\left(h_{i}(\boldsymbol{z},u)-\dot{z}_{i}\right)dt,\label{eq:AugJ_def}
\end{align}
where $\nu$ and $\boldsymbol{\lambda}=(\lambda_{1},\lambda_{2},...,\lambda_{M})$
are Lagrange multipliers that force the constraints. In Eq.~\eqref{eq:AugJ_def},
the exact energetic cost $\Pi$ in $R$ is replaced by its quadratic
approximation $\widetilde{\Pi}$ (we set $q=1$, because the scaling
of $q$ is offset by $w$). Using the calculus of variations \cite{Kamien:2012:DynOpt,Hull:2013:Optimal},
finding an optimal signal $u(t)$ is reduced to solving the differential
equations given by Eq.~\eqref{eq:ODE_def} and 
\begin{align}
\dot{\lambda}_{i}(t) & =-\frac{\partial}{\partial z_{i}}H(\boldsymbol{z},u,\boldsymbol{\lambda}),\label{eq:lambda_ODE}\\
0 & =\frac{\partial}{\partial u}H(\boldsymbol{z},u,\boldsymbol{\lambda}),\label{eq:u_condition}
\end{align}
where $H(\boldsymbol{z},u,\boldsymbol{\lambda})$ is the Hamiltonian
\cite{Kamien:2012:DynOpt,Hull:2013:Optimal}: 
\[
H(\boldsymbol{z},u,\boldsymbol{\lambda})=w\widetilde{\mathcal{P}}+\sum_{i=1}^{M}\lambda_{i}h_{i}(\boldsymbol{z},u).
\]
We assume vanishing initial values for all moments: $\mu_{i}(0)=0$,
$\gamma_{i}(0)=0$, and $\rho_{ij}(0)=0$ {[}i.e., $z_{i}(0)=0$ for
all $i${]}. For the boundary conditions, $\lambda_{N+i_{\mathrm{out}}}(T)=1$
and $\lambda_{i}(T)=0$ ($i\ne i_{\mathrm{out}},N+i_{\mathrm{out}}$)
are required from the optimal control theory, and $z_{i_{\mathrm{out}}}(T)=\mu_{i_{\mathrm{out}}}^{\mathrm{trg}}$
for the final value of $z_{i}$. There are boundary conditions at
both $t=0$ and $t=T$; this two-point boundary value problem can
be solved numerically by using general solvers (see the supplementary
material).

\section{Results}

We consider the following activation-deactivation decoding motif (Fig.~\ref{fig:model}(b)):
an inactive molecule $\overline{X}_{1}$ is activated to become $X_{1}$,
where the activation is dependent on the input molecule $U$. The
rate equation is $\dot{x}_{1}=\alpha_{1}u\left(x_{1}^{\mathrm{tot}}-x_{1}\right)-\beta_{1}x_{1}$,
where $x_{1}^{\mathrm{tot}}$ is the total concentration $x_{1}^{\mathrm{tot}}=x_{1}+\overline{x}_{1}$
which does not change with time ($\overline{x}_{1}$ is the concentration
of $\overline{X}_{1}$), and $\alpha_{1}$ and $\beta_{1}$ are activation
and deactivation rates, respectively. The activated molecule $X_{1}$
activates an output molecule $X_{2}$ (e.g. an activated transcription
factor), and hence $X_{2}$ reports the result (i.e., $i_{\mathrm{out}}=2$).
The rate equation is $\dot{x}_{2}(t)=\alpha_{2}^{0}x_{1}(x_{2}^{\mathrm{tot}}-x_{2})-\beta_{2}x_{2}$,
where $x_{2}^{\mathrm{tot}}=x_{2}+\overline{x}_{2}$, and $\alpha_{2}^{0}$
and $\beta_{2}$ are activation and deactivation rates, respectively
($\overline{x}_{2}$ is the concentration of $\overline{X}_{2}$).
When $x_{2}$ is far from its saturation concentration, activation
of $X_{2}$ is approximately the first-order reaction with respect
to $X_{1}$, i.e., $\dot{x}_{2}=\alpha_{2}x_{1}-\beta_{2}x_{2}$ where
$\alpha_{2}=\alpha_{2}^{0}x_{2}^{\mathrm{tot}}$. Then dynamics of
$X_{1}$ and $X_{2}$ is represented by the following linear differential
equation:
\begin{align}
\dot{x}_{1}(t) & =\alpha_{1}u\left(x_{1}^{\mathrm{tot}}-x_{1}\right)-\beta_{1}x_{1},\label{eq:ODE_1}\\
\dot{x}_{2}(t) & =\alpha_{2}x_{1}-\beta_{2}x_{2}.\label{eq:ODE_2}
\end{align}
Because of the linearity of Eqs.~\eqref{eq:ODE_1} and \eqref{eq:ODE_2},
the moment equation can be obtained without the truncation approximation
\footnote{Note that linearity is not a prerequisite for applying the moment
method. For nonlinear cases, closed moment equations can be obtained
by truncating higher order moments than the second. }. As denoted above, the input signal $u(t)$ must produce the dynamics
that satisfy the constraint that the target mean concentration of
$X_{2}$ at time $t=T$ is $\mu_{2}^{\mathrm{trg}}$ ($i_{\mathrm{out}}=2$
in Eq.~\eqref{eq:endpoint_constraint}). This type of decoding motif
is prevalent and can be found in various biochemical systems \cite{Sasagawa:2005:ERK,Salazar:2008:Decoding,Tanase:2006:NoiseBiol,Hansen:2013:DynDec}.
We call this a \emph{two-stage model}. By incorporating intrinsic
noise due to a small number of molecules, we have a corresponding
FPE from Eq.~\eqref{eq:FPE_def} (see the supplementary material).
We then calculate the moment equation from the FPE. 

\begin{figure*}
\begin{centering}
\includegraphics[width=16cm]{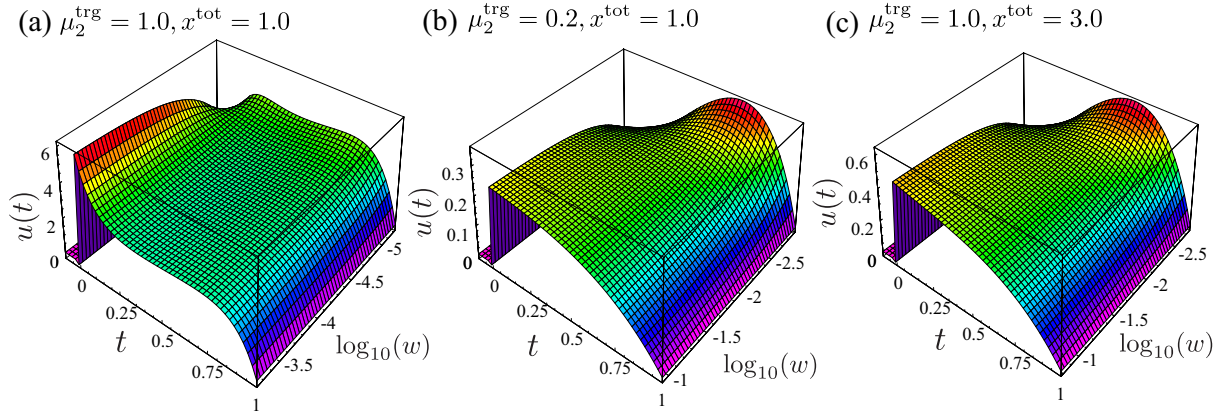}
\par\end{centering}

\protect\caption{Optimal signal $u(t)$ for the two-stage model as a function of $t$
and $\log_{10}(w)$ for three cases: (a) $\mu_{2}^{\mathrm{trg}}=1.0$
and $x_{1}^{\mathrm{tot}}=1.0$, (b) $\mu_{2}^{\mathrm{trg}}=0.2$
and $x_{1}^{\mathrm{tot}}=1.0$, and (c) $\mu_{2}^{\mathrm{trg}}=1.0$
and $x_{1}^{\mathrm{tot}}=3.0$. The range of $w$ is $6.0\times10^{-6}\le w\le6.0\times10^{-4}$
($-5.22\le\log_{10}w\le-3.22$) for (a), $1.2\times10^{-3}\le w\le1.2\times10^{-1}$
($-2.92\le\log_{10}w\le-0.92$) for (b), and $1.7\times10^{-3}\le w\le1.7\times10^{-1}$
($-2.77\le\log_{10}w\le-0.77$) for (c). The other parameters are
$\alpha_{1}=2.0$, $\alpha_{2}=2.0$, $\beta_{1}=1.0$, $\beta_{2}=1.0$,
$Q=0.001$, and $T=1.0$. \label{fig:3DPlot1}}
\end{figure*}

\subsection{Steep pattern minimizes energetic cost, and gradual pattern minimizes
uncertainty}

Using the two-stage model, we calculated the optimal signal $u(t)$.
Figures~\ref{fig:3DPlot1}(a)--(c) shows the optimal signal $u(t)$
as a function of $t$ and $\log_{10}(w)$ for three settings: (a)
$\mu_{2}^{\mathrm{trg}}=1.0$ and $x_{1}^{\mathrm{tot}}=1.0$, (b)
$\mu_{2}^{\mathrm{trg}}=0.2$ and $x_{1}^{\mathrm{tot}}=1.0$, and
(c) $\mu_{2}^{\mathrm{trg}}=1.0$ and $x_{1}^{\mathrm{tot}}=3.0$.
Note that $\mu_{2}^{\mathrm{trg}}=1.0$ is a near-saturation value
with $x_{1}^{\mathrm{tot}}=1.0$, i.e., it is close to the maximal
reachable target concentration. The other parameters are shown in
the caption of Fig.~\ref{fig:3DPlot1}. The minimum value for $w$
is determined such that $u(t)$ satisfies $u(t)\ge0$, and the maximum
is $10^{2}$ times the minimum \footnote{When $w$ is below the minimal values, the signal determined by the
optimal control approach violates the positivity condition at an earlier
time. For such values, the optimal pattern is a very low concentration
at an earlier time and a peak concentration at a later time. }. In all the three cases {[}Figs.~\ref{fig:3DPlot1}(a)--(c){]},
for larger values of $w$, we see that the optimal signals steeply
increase at $t=0$ and gradually decay as time elapses. As $w$ decreases,
the optimal pattern varies from steeper to more-gradual patterns.
Comparing Figs.~\ref{fig:3DPlot1}(a) and (b), we can see the effect
of different target concentrations. The target concentration for Fig.~\ref{fig:3DPlot1}(b)
is lower than it is for (a) (the other parameters are the same). For
Fig.~\ref{fig:3DPlot1}(a), the decay right after $t=0$ is especially
rapid, and this is followed by a plateau state ($t=0.2$--$0.8$);
this is a typical overshooting pattern, similar to that shown in Fig.~\ref{fig:Phos_U}(c).
However, the optimal signal of Fig.~\ref{fig:3DPlot1}(b) does not
exhibit overshooting. We next compared Figs.~\ref{fig:3DPlot1}(a)
and (c) when all parameters are identical except for $x_{1}^{\mathrm{tot}}$
{[}$x_{1}^{\mathrm{tot}}$ for Fig.~\ref{fig:3DPlot1}(c) is larger
than it is for (a){]} to see how the overshooting pattern depends
on the total concentration $x_{1}^{\mathrm{tot}}$. When the total
concentration $x_{1}^{\mathrm{tot}}$ is larger, the optimal signal
does not overshoot. From these results, we see that the steep pattern
minimizes the energy, whereas the gradual pattern minimizes the variance.
Along with the condition of the steep pattern, the overshooting pattern
emerges when the target concentration $\mu_{2}^{\mathrm{trg}}$ is
relatively high and the total concentration $x_{1}^{\mathrm{tot}}$
is relatively low. 

\begin{figure}
\begin{centering}
\includegraphics[width=12cm]{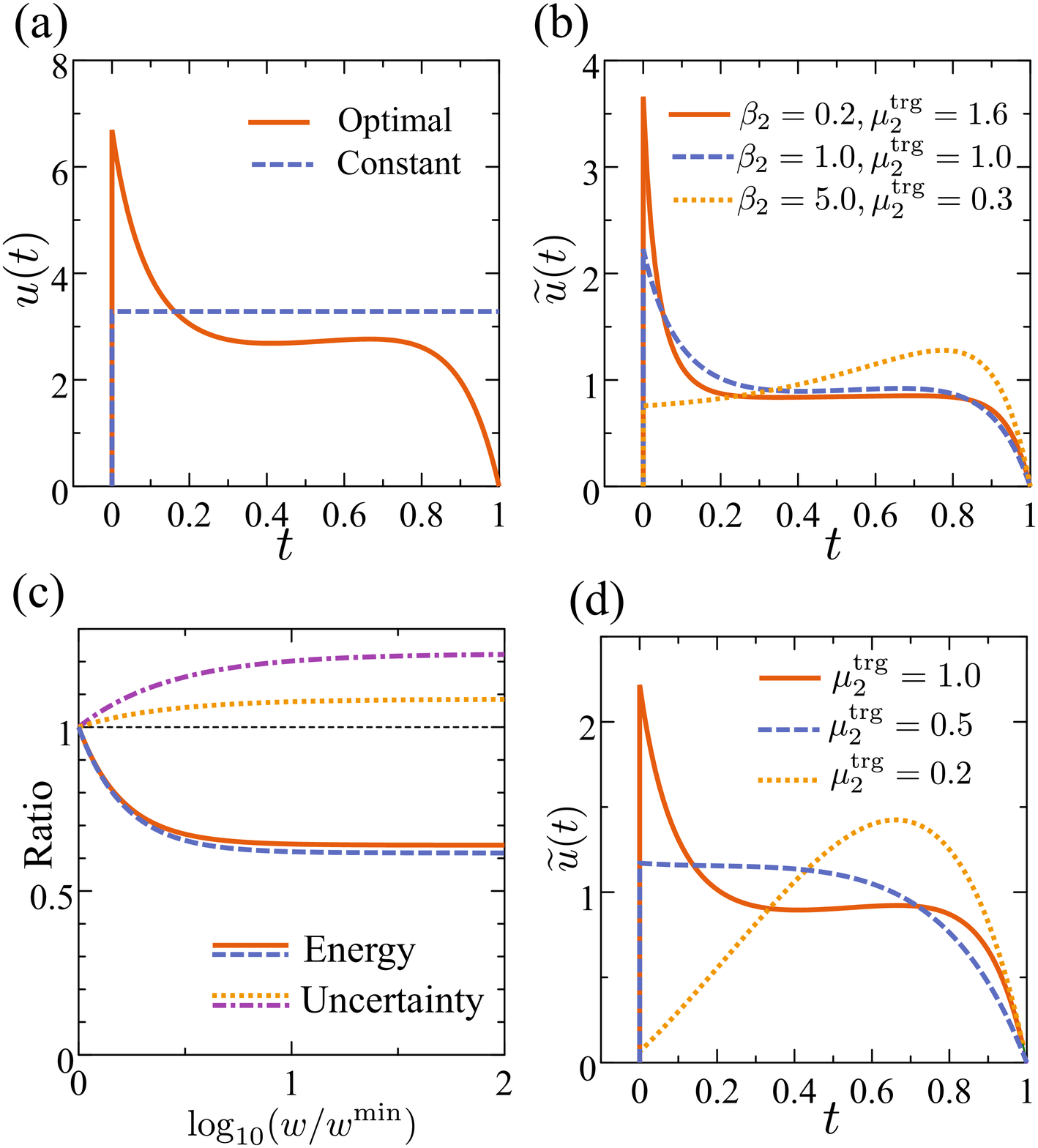}
\par\end{centering}

\protect\caption{Results of the two-stage model. (a) Comparison of the optimal signal
(solid line) and a constant signal (dashed line), which attain the
same target concentration $\mu_{2}^{\mathrm{trg}}=1.0$ ($w=1.0$
for the optimal signal). (b) Normalized signal $\widetilde{u}(t)$
for three values of $\beta_{1}$: $\beta_{2}=0.2$ and $\mu_{2}^{\mathrm{trg}}=1.6$
(solid line), $\beta_{2}=1.0$ and $\mu_{2}^{\mathrm{trg}}=1.0$ (dashed
line), and $\beta_{2}=5.0$ and $\mu_{2}^{\mathrm{trg}}=0.3$ (dotted
line). We set $w=1.0$. (c) Energy and uncertainty dependence on $w$
for two cases: $\mu_{2}^{\mathrm{trg}}=0.2$, $\alpha_{1}=2.0$, $\alpha_{2}=2.0$,
$\beta_{1}=1.0$, and $\beta_{2}=1.0$ {[}parameters are the same
as those in Fig.~\ref{fig:3DPlot1}(b); solid (energy) and dotted
(uncertainty) lines{]} and $\mu_{2}^{\mathrm{target}}=0.5$, $\alpha_{1}=0.5$,
$\alpha_{2}=10.0$, $\beta_{1}=0.1$, and $\beta_{2}=0.1$ {[}dashed
(energy) and dot-dashed (uncertainty) lines{]}. Ranges of $w$ are
$1.2\times10^{-3}\le w\le1.2\times10^{-1}$ for the former and $0.03\le w\le3.0$
for the latter. We plot the ratio of energy and uncertainty to those
at the minimum weight $w^{\min}$. (d) Variation of the normalized
signal $\widetilde{u}(t)$ in response to the target concentration
$\mu_{2}^{\mathrm{trg}}$ for $w=1.2\times10^{-3}$: $\mu_{2}^{\mathrm{trg}}=0.2$
(dotted line), $0.5$ (dashed line), and $1.0$ (solid line). In (a)--(d),
the unspecified parameters are the same as those in Fig.~\ref{fig:3DPlot1}(a).
\label{fig:var_Y}}
\end{figure}

In Fig.~\ref{fig:var_Y}(a), we compared the optimal signal (solid
line), which exhibits the overshooting pattern, with a constant signal
(dashed line), for the period starting at $t=0$ and ending at $t=T$;
both patterns attain the same target concentration $\mu_{2}^{\mathrm{trg}}=1.0$
($w=1.0$ for the optimal signal, which is large enough to show overshooting;
the other parameters are the same as in Fig.~\ref{fig:3DPlot1}(a)).
Although in the interval $t=0$--$0.2$, the concentration of the
optimal signal is larger than that of the constant signal, the optimal
one yields a smaller concentration for $t>0.2$. The energy (quadratic
approximation) of the optimal signal is $\widetilde{\Pi}_{\mathrm{opt}}=9.02$
whereas that of the constant one is $\widetilde{\Pi}_{\mathrm{const}}=10.77$
and thus the ratio is $\widetilde{\Pi}_{\mathrm{opt}}/\widetilde{\Pi}_{\mathrm{const}}=0.84$.
We also calculated the ratio for the exact energy definition $\Pi$
(parameter details are the same as in Fig.~\ref{fig:exact_vs_quadratic}(a))
and we obtained $\Pi_{\mathrm{opt}}/\Pi_{\mathrm{const}}=0.83$\footnote{Because the exact energetic cost diverges to $\infty$ for $u\rightarrow u^{\min}$,
$u(t)$ is truncated at $u^{\mathrm{zf}}$ (the concentration lower
than $u^{\mathrm{zf}}$ is identified as $u^{\mathrm{zf}}$). }, where $\Pi_{\mathrm{opt}}$ and $\Pi_{\mathrm{const}}$ are defined
analogously. Therefore the optimal signal obtained by the quadratic
approximation is energetically efficient with the exact definition. 

\subsection{Slow relaxation of the decoder causes overshooting}

We considered the effects of time scale of the decoder on overthooting
of the optimal signal. When increasing (decreasing) $\beta_{1}$ or
$\beta_{2}$, the relaxation time of the decoder becomes shorter (longer).
We first varied $\beta_{2}$ while keeping other parameters unchanged,
except for $\mu_{2}^{\mathrm{trg}}$ (the other parameters are the
same as those in Fig.~\ref{fig:var_Y}(a)). Since we are interested
in the shape rather than the magnitude, we normalized the signal as
follows: $\widetilde{u}(t)=u(t)/\sqrt{\widetilde{\Pi}}$, which guarantees
the unit energy $\widetilde{\Pi}=1$. Figure~\ref{fig:var_Y}(b)
shows the normalized signal $\widetilde{u}(t)$ for three cases: $\beta_{2}=0.2$
(solid line), $\beta_{2}=1.0$ (dashed line), and $\beta_{2}=5.0$
(dotted line). As shown above, the overshooting pattern emerges when
$\mu_{2}^{\mathrm{trg}}$ is close to the saturation value. Thus,
for each $\beta_{2}$ value, $\mu_{2}^{\mathrm{trg}}$ is set to near
the maximal reachable value (parameter details are shown in the caption
of Fig~\ref{fig:var_Y}(b)). When $\beta_{2}=5.0$, the optimal signal
does not overshoot, but the other two settings do. From this result,
when the relaxation time of $X_{2}$ is sufficiently shorter than
$T$, the steep (overshooting) pattern does not minimize energy consumption.
The same calculation was performed for $\beta_{1}$ and we found that
all of the optimal signals exhibit the overshooting (see the supplementary
material). This implies that the relaxation time of $X_{1}$ is not
responsible for the overshooting pattern.

\subsection{Tradeoff between energy and uncertainty}

In Fig.~\ref{fig:var_Y}(c), we next evaluated the dependence on
the weight $w$ of the energy $\widetilde{\Pi}$ and the uncertainty
$\gamma_{i_{\mathrm{out}}}(T)$ for two parameter settings. Parameters
of the first setting is the same as those in Fig.~\ref{fig:3DPlot1}(b)
whose results are plotted by solid (energy) and dotted (uncertainty)
lines. The second setting highlights the uncertainty variation when
we change $w$ and these results are shown by dashed (energy) and
dot-dashed (uncertainty) lines (parameter details are described in
the caption of Fig.~\ref{fig:var_Y}(c)). In Fig.~\ref{fig:var_Y}(c),
we plot the ratio of the energy and uncertainty to those at the minimum
weight $w^{\min}$, where $w^{\min}$ is the minimum of $w$ for each
parameter setting. For the both settings, as $w$ increases, the uncertainty
increases, and the energy decreases; there is thus a tradeoff between
the energy and the uncertainty, since they cannot be minimized simultaneously.
For cellular inference, the tradeoff between uncertainty and energy
consumption has been confirmed by several studies \cite{Tu:2008:MaxwellDemon,Mehta:2012:Energetic,Lang:2014:InferCell,Barato:2014:CellInfo}.
Similarly, in biochemical clocks, it has been shown that there is
a tradeoff between temporal accuracy and energy consumption \cite{Cao:2015:ClockEnergy}.
These studies \cite{Mehta:2012:Energetic,Lang:2014:InferCell,Barato:2014:CellInfo,Cao:2015:ClockEnergy}
calculated the entropy production, which is the energy required for
maintaining a system at NESS. There is also a tradeoff between the
energy cost and information coding in neural systems (\cite{Sengupta:2014:PC}
and references therein). We have shown that a similar relation also
holds for dynamical signals.

\subsection{Calculation with simplified model}

We identified that the steep (overshooting) pattern minimizes the
energy. Let us explain this mechanism with a simplified two-stage
model: 
\begin{equation}
\dot{x}_{1}(t)=u(t),\hspace*{1em}\dot{x}_{2}(t)=\alpha_{2}x_{1}-\beta_{2}x_{2},\label{eq:simplified_model}
\end{equation}
along with a delta-function stimulus, $u(t)=\delta(t-t_{s})$ {[}$t_{s}$
is the time of the stimulus, where $0<t_{s}<T${]}. From Eq.~\eqref{eq:simplified_model},
the final output concentration is 
\begin{equation}
x_{2}(T)=\frac{\alpha_{2}}{\beta_{2}}\left\{ 1-e^{-\beta_{2}\left(T-t_{s}\right)}\right\} .\label{eq:sol_simplified_model}
\end{equation}
When the relaxation of $X_{2}$ is very slow ($\beta_{2}^{-1}\gg T$),
the output concentration is $x_{2}(T)\simeq\alpha_{2}(T-t_{s})$,
which shows that the signal at an earlier time has a greater effect
than at a later time; the steep pattern allows the output concentration
to reach the target concentration at a lower cost. For the fast relaxation
case ($\beta_{2}^{-1}\ll T$), we obtain $x_{2}(T)\simeq\alpha_{2}\beta_{2}^{-1}$,
which shows that $x_{2}(T)$ does not depend on $t_{s}$; it is not
advantageous for the signal to have a peak at an earlier time. Therefore,
when the relaxation time of the decoder is very fast, the steep pattern
does not confer advantages for minimizing the energy consumption;
this agrees with the optimal control calculation shown in Fig.~\ref{fig:var_Y}(b).
Next, suppose there is only a single stage required to decode the
signal $\dot{x}_{1}(t)=u(t)$, where $x_{1}(T)$ is the output concentration
(a single-stage model). In this case, the final output concentration
is $x_{1}(T)=1$ and does not depend on $t_{s}$, implying that the
steep patterns do not minimize the energy. On the other hand, when
there is a third stage $\dot{x}_{3}=\alpha_{3}x_{2}-\beta_{3}x_{3}$
in addition to Eqs.~\eqref{eq:simplified_model} (a three-stage model),
we find that the final concentration $x_{3}(T)$ depends on $t_{s}$,
unless $\beta_{2}^{-1},\beta_{3}^{-1}\ll T$ {[}for $\beta_{2}^{-1},\beta_{3}^{-1}\gg T$,
we have $x_{3}(T)\simeq\alpha_{2}\alpha_{3}(T-t_{s})^{2}/2$, and
for $\beta_{2}^{-1},\beta_{3}^{-1}\ll T$, $x_{3}(T)=\alpha_{2}\alpha_{3}/(\beta_{2}\beta_{3})${]}.
Thus, the steep patterns generally minimize the energy consumption
when the relaxation of the decoder is slow and there are more than
two stages in the decoding. 

The effect of the gradual pattern can be accounted for by the moment
equation. The variance $\gamma_{2}$ and covariance $\rho_{12}$ are
governed by (cf. moment equations in the supplementary material) 
\begin{align}
\dot{\gamma}_{2}(t) & =2\left(\alpha_{2}\rho_{12}-\beta_{2}\gamma_{2}\right)+2Q\left(\alpha_{2}\mu_{1}+\beta_{2}\mu_{2}\right),\label{eq:gamma2_def}\\
\dot{\rho}_{12}(t) & =\alpha_{2}\gamma_{1}-\rho_{12}\left(\alpha_{1}u+\beta_{1}+\beta_{2}\right).\label{eq:rho_def}
\end{align}
We find that the main reason for the difference between the variance
$\gamma_{2}(T)$ of the steep and gradual patterns is the area $\int_{0}^{T}\rho_{12}(t)dt$;
namely, a smaller area yields a smaller value for $\gamma_{2}(T)$.
From Eq.~\eqref{eq:rho_def}, because the decay velocity of $\rho_{12}(t)$
depends on $u$, a higher concentration of $u$ at a later time results
in a smaller value for $\rho_{12}(t)$, which corresponds to the gradual
pattern of $u(t)$.

\section{Discussion and Conclusion}

Our result provides insights into experimentally observed dynamical
patterns. Reference~\cite{Selimkhanov:2014:DynSig} reported the
dynamical pattern of ERK activity in response to different strengths
of extracellular stimulus (i.e., the ligand concentration); the pattern
is steep when stimulated by a strong stimulus, and it is gradual when
stimulated by a weak one. This experimental observation can be accounted
for by our model. We show that the optimal signal is steep for larger
values of $\mu_{2}^{\mathrm{trg}}$ and gradual for smaller values
(Fig.~\ref{fig:var_Y}(d)). It is expected that the strong and weak
ligand stimuli result in strong and weak responses, respectively,
i.e., higher and lower output concentrations. Therefore, the ERK activity
induced by the strong ligand stimulus may be related to $\mu_{2}^{\mathrm{trg}}=1.0$,
and that induced by the weak one is related to $\mu_{2}^{\mathrm{trg}}=0.2$.
When the target concentration is higher (i.e., $\mu_{2}^{\mathrm{trg}}=1.0$),
the magnitude of the signal is larger, and hence the effect of the
energy of the signal on the objective function $R$ (Eq.~\eqref{eq:J_def1})
is greater than that of the variance $\gamma_{2}(T)$. In contrast,
for the smaller values of $\mu_{2}^{\mathrm{trg}}$ (i.e., $\mu_{2}^{\mathrm{trg}}=0.2$),
the variance $\gamma_{2}(T)$ becomes the leading term because the
energy of the signal is smaller. Therefore, the steep pattern is preferable
when the target concentration is higher, while the gradual one is
preferable when the target concentration is lower. These theoretical
results qualitatively agree with the observed dynamical patterns reported
in Ref.~\cite{Selimkhanov:2014:DynSig}.

Along with conditions for the steep pattern, the overshooting dynamics
minimize the energy of the input signals when the total concentration
$x_{1}^{\mathrm{tot}}$ is smaller and the target concentration $\mu_{i_{\mathrm{out}}}^{\mathrm{trg}}$
is higher. Surprisingly, this behavior can be found in several dynamical
patterns; for example, activities of the ERK, the I$\kappa$B kinase
(IKK), which regulates the transcription factor NF-$\kappa$B, and
the kinase AKT show this behavior \cite{Werner:2005:IKK,Sasagawa:2005:ERK,Werner:2008:IKK,Kubota:2012:InsulinAKT,Mathew:2014:PI3KAKT}.
These examples indicate that the pattern has biological advantages.
We also note the biochemical origin of the overshoot. For example,
simple incoherent feed-forward loops \cite{Mangan:2003:FFL,Mangan:2006:FFL,Alon:2007:NetMotif,Alon:2007:SystBiolBook}
and activation-deactivation motifs \cite{Behar:2013:Tunable} can
generate such a pattern, and these motifs can indeed be found in signaling
pathways. Furthermore, a strongly damped oscillation is indistinguishable
from overshooting. Although NF-$\kappa$B is known to exhibit damped
oscillation upon stimulation, some studies \cite{Barken:2005:Comment,Cheong:2008:NFkBreview}
are skeptical about the functional role of the NF-$\kappa$B oscillation;
that is, the NF-$\kappa$B oscillation may be a by-product of inducing
overshooting. Overshooting has often been observed in actual dynamical
patterns, but its functional advantage has not been well understood.
We have shown that this pattern produces direct benefits.

\section*{Acknowledgments}

This work was supported by KAKENHI Grant No.~16K00325 from the Ministry
of Education, Culture, Sports, Science and Technology.

\providecommand{\newblock}{}

\end{document}


\title{Supplementary Material for\\
 ``Optimal Temporal Patterns for Dynamical Cellular Signaling''}

\author{Yoshihiko Hasegawa}

\maketitle
This supplementary material describes in detail the calculations introduced
in the main text. Equation and figure numbers in this section are
prefixed with S (e.g., Eq.\textbf{~}(S1) or Fig.~S1). Numbers without
the prefix (e.g., Eq.~(1)or Fig.~1) refer to items in the main text.

\section{Energetic cost of signal pattern}

We derive the energetic cost of signal $U$ with a simple biochemical
model. Details of the thermodynamics of biochemical reactions can
be found in Refs.~\cite{Qian:2005:Thermodynamics,Qian:2007:PiReview,Beard:2008:ChemicalBook}.
$U$ is activated from $\overline{U}$ by the following reaction:
\begin{equation}
\overline{U}\xrightleftharpoons[\beta_{u}]{\alpha_{u}}U,\label{eq:U_reaction0}
\end{equation}
where $\alpha_{u}$ and $\beta_{u}$ are reaction rates. The time
evolution of the concentration $u$ and $\overline{u}$ is
\begin{equation}
\frac{d\overline{u}}{dt}=-\frac{du}{dt}=\beta_{u}u-\alpha_{u}\overline{u},\label{eq:ODE0}
\end{equation}
where $u$ and $\overline{u}$ are the concentrations of $U$ and
$\overline{U}$, respectively (the total concentration $u^{\mathrm{tot}}=u+\overline{u}$
does not change with time). As described in the main text, the deactivation
reaction is not the reverse reaction of activation. The activation
and deactivation of $U$ are mediated by phosphorylation and dephosphorylation,
respectively. These reactions are written as (cf. Eq.~\UUreaction)
\begin{align}
\overline{U}+\mathrm{ATP} & \xrightleftharpoons[\beta_{a}^{0}]{\alpha_{a}^{0}}U+\mathrm{ADP},\label{eq:U_reaction1}\\
U & \xrightleftharpoons[\beta_{d}^{0}]{\alpha_{d}}\overline{U}+\mathrm{P}_{i},\label{eq:U_reaction2}
\end{align}
where ATP, ADP, and $\mathrm{P}_{i}$ are adenosine triphosphate,
adenosine diphosphate, and inorganic phosphate, respectively; and
$\alpha_{a}^{0}$, $\beta_{a}^{0}$, $\alpha_{d}$, and $\beta_{d}^{0}$
are reaction rates. According to the mass action kinetics, the time
evolution of the concentrations $u$ and $\overline{u}$ are
\begin{equation}
\frac{d\overline{u}}{dt}=-\frac{du}{dt}=\beta_{a}^{0}c_{D}u-\beta_{d}^{0}c_{P}\overline{u}+\alpha_{d}u-\alpha_{a}^{0}c_{T}\overline{u},\label{eq:ODE}
\end{equation}
where $c_{T}$, $c_{D}$, and $c_{P}$ are the concentrations of ATP,
ADP, and $\mathrm{P}_{i}$, respectively. When reaction rates ($\alpha_{a}^{0}$,
$\beta_{a}^{0}$, $\alpha_{d}$, and $\beta_{d}^{0}$) are held fixed
and the system is closed, the system relaxes to an equilibrium state.
However, in cellular environments where the system is open, the concentrations
of ATP, ADP, and $\mathrm{P}_{i}$ are approximately constant due
to external agents, as described in the main text (we denote the cellular
concentrations as $c_{T}=c_{T}^{0}$, $c_{D}=c_{D}^{0}$, and $c_{P}=c_{P}^{0}$).
Therefore, the system relaxes to a nonequilibrium steady state (NESS).
From Eq.~\eqref{eq:ODE}, the steady-state concentration $u^{\mathrm{ss}}$
is
\begin{equation}
u^{\mathrm{ss}}=\frac{u^{\mathrm{tot}}(\beta_{d}+\alpha_{a})}{\beta_{a}+\beta_{d}+\alpha_{a}+\alpha_{d}}.\label{eq:uss_def}
\end{equation}
where $\alpha_{a}=\alpha_{a}^{0}c_{T}^{0}$, $\beta_{a}=\beta_{a}^{0}c_{D}^{0}$,
and $\beta_{d}=\beta_{d}^{0}c_{P}^{0}$ (note that $\alpha_{u}=\alpha_{a}+\beta_{d}$
and $\beta_{u}=\alpha_{d}+\beta_{a}$). The steady-state net flux
(clockwise direction in Fig.~\FIGPhosUU(d)~in the main text) is
\begin{equation}
\mathcal{J}^{\mathrm{ss}}=\frac{u^{\mathrm{tot}}\left(\alpha_{a}\alpha_{d}-\beta_{a}\beta_{d}\right)}{\beta_{a}+\beta_{d}+\alpha_{a}+\alpha_{d}}.\label{eq:Jss_def}
\end{equation}
The free-energy is dissipated along a cycle $\overline{U}\rightarrow U\rightarrow\overline{U}$
(the clockwise direction in Fig.~\FIGPhosUU(d)). The chemical potential
difference of the cycle is 
\begin{equation}
\Delta\phi=\Delta\phi_{1}+\Delta\phi_{2},\label{eq:Delta_mu_def}
\end{equation}
where $\Delta\phi_{1}$ and $\Delta\phi_{2}$ are chemical potential
differences of Eqs.~\eqref{eq:U_reaction1} and \eqref{eq:U_reaction2},
respectively:
\begin{equation}
\Delta\phi_{1}=\Delta\phi_{1}^{0}+k_{B}\mathcal{T}\ln\left(\frac{uc_{D}^{0}}{\overline{u}c_{T}^{0}}\right),\hspace*{1em}\Delta\phi_{2}=\Delta\phi_{2}^{0}+k_{B}\mathcal{T}\ln\left(\frac{\overline{u}c_{P}^{0}}{u}\right),\label{eq:delta_mu_def}
\end{equation}
where $\Delta\phi_{1}^{0}$ and $\Delta\phi_{2}^{0}$ are the chemical
potentials of the standard state,  $k_{B}$ is the Boltzmann constant,
and $\mathcal{T}$ is the temperature. Because each of the chemical
potential differences is zero at equilibrium, that is $\Delta\phi_{1}=\Delta\phi_{2}=0$,
we obtain
\begin{equation}
\Delta\phi_{1}^{0}=-k_{B}\mathcal{T}\ln\left(\frac{u^{\mathrm{eq}}c_{D}^{\mathrm{eq}}}{\overline{u}^{\mathrm{eq}}c_{T}^{\mathrm{eq}}}\right),\hspace*{1em}\Delta\phi_{2}^{0}=-k_{B}\mathcal{T}\ln\left(\frac{\overline{u}^{\mathrm{eq}}c_{P}^{\mathrm{eq}}}{u^{\mathrm{eq}}}\right),\label{eq:Eq_conditions}
\end{equation}
where the superscript ``eq'' denotes the concentration at equilibrium.
Furthermore, the net flux $\mathcal{J}$ vanishes at equilibrium (detailed
balance), which yields the following condition:
\begin{equation}
\alpha_{a}^{0}\alpha_{d}c_{T}^{\mathrm{eq}}-\beta_{a}^{0}\beta_{d}^{0}c_{D}^{\mathrm{eq}}c_{P}^{\mathrm{eq}}=0.\label{eq:Jeq_def}
\end{equation}
From Eqs.~\eqref{eq:Eq_conditions} and \eqref{eq:Jeq_def}, the
free energy dissipation of a single cycle (Eq.~\eqref{eq:Delta_mu_def})
is
\begin{equation}
\Delta\phi=k_{B}\mathcal{T}\ln\left(\frac{\beta_{a}^{0}\beta_{d}^{0}c_{D}^{0}c_{P}^{0}}{\alpha_{a}^{0}\alpha_{d}c_{T}^{0}}\right)=k_{B}\mathcal{T}\ln\left(\frac{\beta_{a}\beta_{d}}{\alpha_{a}\alpha_{d}}\right).\label{eq:chempot_onecycle}
\end{equation}
From Eqs.~\eqref{eq:Jss_def} and \eqref{eq:chempot_onecycle}, the
instantaneous free energy dissipation $\mathcal{P}$ (i.e. power),
which is an analogue of $[\mathrm{power}]=[\mathrm{current}]\times[\mathrm{volage}]$
of electric circuits, is 
\begin{equation}
\mathcal{P}=-\mathcal{J}^{\mathrm{ss}}\Delta\phi=k_{B}\mathcal{T}\frac{u^{\mathrm{tot}}\left(\alpha_{a}\alpha_{d}-\beta_{a}\beta_{d}\right)}{\beta_{a}+\beta_{d}+\alpha_{a}+\alpha_{d}}\ln\left(\frac{\alpha_{a}\alpha_{d}}{\beta_{a}\beta_{d}}\right).\label{eq:dissipation}
\end{equation}
From Eq.~\eqref{eq:dissipation}, the energy dissipated during the
period $0\le t\le T$ is
\begin{equation}
\Pi=\int_{0}^{T}\mathcal{P}dt.\label{eq:Pi_def_Supp}
\end{equation}
We assume that kinase activity is controlled by an upstream molecular
species and thus $\alpha_{a}^{0}$ (i.e. $\alpha_{a}$) varies temporally.
When relaxation of the system to NESS is sufficiently fast {[}i.e.,
$1/(\beta_{a}+\beta_{d}+\alpha_{a}+\alpha_{d})\ll T${]}, the concentration
$u$ is well approximated by $u^{\mathrm{ss}}$ for the time-varying
case. From Eq.~\eqref{eq:uss_def}, we can write $\alpha_{a}$ in
terms of $u$ as follows: 
\begin{equation}
\alpha_{a}=\frac{u}{u^{\mathrm{tot}}-u}\left(\alpha_{d}+\beta_{a}\right)-\beta_{d}.\label{eq:alpha_a_def}
\end{equation}
Because $\alpha_{a}>0$, $u$ must satisfy $u>u^{\min}$, where 
\begin{equation}
u^{\min}=\frac{u^{\mathrm{tot}}\beta_{d}}{\beta_{a}+\beta_{d}+\alpha_{d}}.\label{eq:ucond_1}
\end{equation}
Because it is numerically difficult to use Eq.~\eqref{eq:dissipation}
in the optimal control calculations, we approximate the power $\mathcal{P}$
with a simple equation. $\mathcal{P}=0$ at a zero-flux point $u=u^{\mathrm{zf}}$
with 
\begin{equation}
u^{\mathrm{zf}}=\frac{u^{\mathrm{tot}}\beta_{d}}{\beta_{d}+\alpha_{d}},\label{eq:ucond_2}
\end{equation}
where $\mathcal{J}$ vanishes (note that $u^{\mathrm{zf}}>u^{\min}$
always holds). Assuming that $u^{\mathrm{zf}}$ is sufficiently small
(which may be approximated as $0$), $\mathcal{P}=0$ when $u$ is
very low concentration. Furthermore, $\mathcal{P}$ increases superlinearly
as $u$ increases from $u=u^{\mathrm{zf}}$. Also the approximation
needs to be computationally feasible for the optimal control. Taking
into account the above requirements, we may approximate $\mathcal{P}\simeq\widetilde{\mathcal{P}}$
with 
\begin{equation}
\widetilde{\mathcal{P}}=qu^{2},\label{eq:pol_approx}
\end{equation}
where $q>0$ is a proportionality coefficient. From Eq.~\eqref{eq:pol_approx},
we represent the energy of the signal as 
\[
\widetilde{\Pi}=\int_{0}^{T}\widetilde{\mathcal{P}}dt=\int_{0}^{T}qu(t)^{2}dt.
\]
In the main calculation, we employ $q=1$, because the scaling of
$q$ is offset by the weight parameter $w$ in the performance index
(cf. Eq.~\JUdefI).

\section{Moment equation}

In this section, we derive the equations that must be satisfied by
the mean, variance, and covariance. We consider the Fokker--Planck
equation (FPE): 
\begin{equation}
\frac{\partial}{\partial t}P(\boldsymbol{x};t)=-\sum_{i=1}^{N}\frac{\partial}{\partial x_{i}}F_{i}(\boldsymbol{x};t)P(\boldsymbol{x};t)+\sum_{i=1}^{N}\frac{\partial^{2}}{\partial x_{i}^{2}}G_{i}(\boldsymbol{x};t)P(\boldsymbol{x};t),\label{eq:FPE_def}
\end{equation}
where $\boldsymbol{x}=(x_{1},x_{2},...,x_{N})$, $P(\boldsymbol{x};t)$
is the probability density of $\boldsymbol{x}$ at time $t$, and
$F_{i}(\boldsymbol{x};t)$ and $G_{i}(\boldsymbol{x};t)$ are the
drift and diffusion terms, respectively (we do not consider cross
terms, such as $\partial^{2}/\partial x_{i}\partial x_{j}$ ($i\ne j$),
as these terms do not emerge in our model). We denote the range of
$x_{i}$ as $x_{i}^{\min}\le x_{i}\le x_{i}^{\max}$; for example,
in the two-stage model, $x_{1}^{\min}=0$ and $x_{1}^{\max}=x_{1}^{\mathrm{tot}}$
for $X_{1}$. Because the concentration must satisfy these constraints,
we impose reflecting walls at the boundaries. Writing the FPE \eqref{eq:FPE_def}
as the continuity equation, we have 
\begin{equation}
\frac{\partial}{\partial t}P(\boldsymbol{x};t)+\sum_{i=1}^{N}\frac{\partial}{\partial x_{i}}J_{i}(\boldsymbol{x};t)=0,\label{eq:continuity_equation}
\end{equation}
where $J_{i}$ denotes the probability current: 
\begin{equation}
J_{i}(\boldsymbol{x};t)=F_{i}(\boldsymbol{x};t)P(\boldsymbol{x};t)-\frac{\partial}{\partial x_{i}}G_{i}(\boldsymbol{x};t)P(\boldsymbol{x};t).\label{eq:prob_current_def}
\end{equation}
Due to the reflecting walls, the current vanishes at the boundaries,
i.e., 
\begin{equation}
J_{i}(\boldsymbol{x};t)=0\hspace*{1em}\mathrm{at}\hspace*{1em}x_{i}=x_{i}^{\min}\,\mathrm{and}\,x_{i}=x_{i}^{\max}.\label{eq:vanish_current}
\end{equation}
Here, we consider the (uncentralized) moment ($k\ne\ell$): 
\begin{equation}
\left\langle x_{k}^{m}x_{\ell}^{n}\right\rangle =\int d\boldsymbol{x}\,x_{k}^{m}x_{\ell}^{n}P(\boldsymbol{x};t),\label{eq:moment_def}
\end{equation}
where 
\[
\int d\boldsymbol{x}=\int_{x_{1}^{\min}}^{x_{1}^{\max}}dx_{1}\int_{x_{2}^{\min}}^{x_{2}^{\max}}dx_{2}\cdots\int_{x_{N}^{\min}}^{x_{N}^{\max}}dx_{N}.
\]
The time evolution of the moment obeys 
\begin{align}
\frac{d}{dt}\left\langle x_{k}^{m}x_{\ell}^{n}\right\rangle  & =\int d\boldsymbol{x}\,x_{k}^{m}x_{\ell}^{n}\frac{\partial}{\partial t}P(\boldsymbol{x};t),\nonumber \\
 & =\sum_{i=1}^{N}\int x_{k}^{m}x_{\ell}^{n}\left[-\frac{\partial}{\partial x_{i}}F_{i}(\boldsymbol{x};t)P(\boldsymbol{x};t)+\frac{\partial^{2}}{\partial x_{i}^{2}}G_{i}(\boldsymbol{x};t)P(\boldsymbol{x};t)\right]d\boldsymbol{x},\label{eq:moment_of_FPE}
\end{align}
where Eq.~\eqref{eq:FPE_def} is used. Using integration by parts,
we have 
\begin{align}
\frac{d}{dt}\left\langle x_{k}^{m}x_{\ell}^{n}\right\rangle  & =-\sum_{i=1}^{N}\int d\boldsymbol{x}_{-i}\left\{ x_{k}^{m}x_{\ell}^{n}J_{i}(\boldsymbol{x};t)\right\} \biggr|_{x_{i}^{\min}}^{x_{i}^{\max}}\nonumber \\
 & +\sum_{i=1}^{N}\int d\boldsymbol{x}\left[\frac{\partial\left(x_{k}^{m}x_{\ell}^{n}\right)}{\partial x_{i}}F_{i}(\boldsymbol{x};t)P(\boldsymbol{x};t)-\frac{\partial\left(x_{k}^{m}x_{\ell}^{n}\right)}{\partial x_{i}}\frac{\partial}{\partial x_{i}}G_{i}(\boldsymbol{x};t)P(\boldsymbol{x};t)\right],\label{eq:MM_mid}
\end{align}
where we formally define 
\[
\int d\boldsymbol{x}_{-i}=\prod_{j=1,j\ne i}^{N}\int_{x_{j}^{\min}}^{x_{j}^{\max}}dx_{j}.
\]
From Eq.~\eqref{eq:vanish_current}, the first term in Eq.~\eqref{eq:MM_mid}
vanishes, and we obtain 
\begin{align*}
\frac{d}{dt}\left\langle x_{k}^{m}x_{\ell}^{n}\right\rangle  & =\sum_{i=1}^{N}\int d\boldsymbol{x}\frac{\partial\left(x_{k}^{m}x_{\ell}^{n}\right)}{\partial x_{i}}F_{i}(\boldsymbol{x};t)P(\boldsymbol{x};t)\\
 & +\sum_{i=1}^{N}\int d\boldsymbol{x}_{-i}\left[-\frac{\partial\left(x_{k}^{m}x_{\ell}^{n}\right)}{\partial x_{i}}G_{i}(\boldsymbol{x};t)P(\boldsymbol{x};t)\biggr|_{x_{i}^{\min}}^{x_{i}^{\max}}+\int dx_{i}\frac{\partial^{2}\left(x_{k}^{m}x_{\ell}^{n}\right)}{\partial x_{i}^{2}}G_{i}(\boldsymbol{x};t)P(\boldsymbol{x};t)\right],
\end{align*}
where we again used integration by parts. If we assume that $G_{i}(\boldsymbol{x};t)P(\boldsymbol{x};t)$
is negligible at the boundaries $x_{i}=x_{i}^{\min}$ and $x_{i}=x_{i}^{\max}$,
we have 
\begin{align}
\frac{d}{dt}\left\langle x_{k}^{m}x_{\ell}^{n}\right\rangle  & =\sum_{i=1}^{N}\left[\left\langle \frac{\partial\left(x_{k}^{m}x_{\ell}^{n}\right)}{\partial x_{i}}F_{i}(\boldsymbol{x};t)\right\rangle +\left\langle \frac{\partial^{2}\left(x_{k}^{m}x_{\ell}^{n}\right)}{\partial x_{i}^{2}}G_{i}(\boldsymbol{x};t)\right\rangle \right].\label{eq:moment_ode}
\end{align}
Equation~\eqref{eq:moment_ode} is an equation for uncentralized
moments. In order to obtain closed equations for the mean, variance,
and covariance for general $F_{i}(\boldsymbol{x};t)$ and $G_{i}(\boldsymbol{x};t)$,
we expand $x_{i}$ around the mean values as $x_{i}-\mu_{i}=\delta x_{i}$,
with $\mu_{i}=\left\langle x_{i}\right\rangle $. Retaining terms
up to the second order, such as $\left\langle \delta x_{k}^{m}\delta x_{\ell}^{n}\right\rangle $
with $m+n=2$, we obtain closed equations with respect to $\mu_{i}(t)$,
$\gamma_{i}(t)=\left\langle \left(x_{i}(t)-\mu_{i}(t)\right)^{2}\right\rangle $,
and $\rho_{ij}=\left\langle \left(x_{i}-\mu_{i}\right)\left(x_{j}-\mu_{j}\right)\right\rangle $.
However note that in the two-stage model, because $F_{i}(\boldsymbol{x};t)$
and $G_{i}(\boldsymbol{x};t)$ are linear with respect to $\boldsymbol{x}$,
we can obtain closed differential equations for $\mu_{i}(t)$, $\gamma_{i}(t)$,
and $\rho_{ij}(t)$ without the truncation.

\section{Differential equations of optimal control\label{sec:Calc_OC}}

Deterministic equations for the two-stage model are given by Eqs.~\ODEUI~and
\ODEUII. From Eq.~\FPEUdef, the corresponding FPE is 
\begin{align}
\frac{\partial}{\partial t}P(\boldsymbol{x};t) & =\left[-\frac{\partial}{\partial x_{1}}\left\{ \alpha_{1}u(x_{1}^{\mathrm{tot}}-x_{1})-\beta_{1}x_{1}\right\} -\frac{\partial}{\partial x_{2}}\left\{ \alpha_{2}x_{1}-\beta_{2}x_{2}\right\} \right.\nonumber \\
 & \left.+Q\frac{\partial^{2}}{\partial x_{1}^{2}}\left\{ \alpha_{1}u(x_{1}^{\mathrm{tot}}-x_{1})+\beta_{1}x_{1}\right\} +Q\frac{\partial^{2}}{\partial x_{2}^{2}}\left\{ \alpha_{2}x_{1}+\beta_{2}x_{2}\right\} \right]P(\boldsymbol{x};t).\label{eq:FPE_2V}
\end{align}
From Eq.~\eqref{eq:moment_ode}, the moment equations are 
\begin{align}
\dot{\mu}_{1} & =\alpha_{1}u(x_{1}^{\mathrm{tot}}-\mu_{1})-\beta_{1}\mu_{1},\label{eq:MM_1}\\
\dot{\mu}_{2} & =\alpha_{2}\mu_{1}-\beta_{2}\mu_{2},\label{eq:MM_2}\\
\dot{\gamma}_{1} & =2Q\left\{ \beta_{1}\mu_{1}+\alpha_{1}u(x_{1}^{\mathrm{tot}}-\mu_{1})\right\} -2\gamma_{1}\left(\alpha_{1}u+\beta_{1}\right),\label{eq:MM_3}\\
\dot{\gamma}_{2} & =2\left(\alpha_{2}\rho_{12}-\beta_{2}\gamma_{2}\right)+2Q\left(\alpha_{2}\mu_{1}+\beta_{2}\mu_{2}\right),\label{eq:MM_4}\\
\dot{\rho}_{12} & =\alpha_{2}\gamma_{1}-\rho_{12}\left(\alpha_{1}u+\beta_{1}+\beta_{2}\right).\label{eq:MM_5}
\end{align}
Differential equations for the Lagrange multiplier $\lambda_{i}$
are obtained from Eq.~\lambdaUODE, as follows: 
\begin{align}
\dot{\lambda}_{1} & =-\alpha_{2}\lambda_{2}-2\lambda_{3}Q(\beta_{1}-\alpha_{1}u)-2\alpha_{2}\lambda_{4}Q+\lambda_{1}(\alpha_{1}u+\beta_{1}),\label{eq:l1_def}\\
\dot{\lambda}_{2} & =\beta_{2}\lambda_{2}-2\beta_{2}\lambda_{4}Q,\label{eq:l2_def}\\
\dot{\lambda}_{3} & =-\alpha_{2}\lambda_{5}+2\lambda_{3}(\alpha_{1}u+\beta_{1}),\label{eq:l3_def}\\
\dot{\lambda}_{4} & =2\beta_{2}\lambda_{4},\label{eq:l4_def}\\
\dot{\lambda}_{5} & =-2\alpha_{2}\lambda_{4}+\lambda_{5}(\beta_{1}+\beta_{2}+\alpha_{1}u).\label{eq:l5_def}
\end{align}
Here, $u(t)$ is obtained from Eq.~\uUcondition: 
\begin{equation}
u(t)=\frac{\alpha_{1}}{2wq}\left(2Q\lambda_{3}\mu_{1}-2Q\lambda_{3}x_{1}^{\mathrm{tot}}+\rho_{12}\lambda_{5}+2\,\gamma_{1}\lambda_{3}+\lambda_{1}\mu_{1}-\lambda_{1}x_{1}^{\mathrm{tot}}\right).\label{eq:u_def}
\end{equation}
According to the optimal control theory \cite{Hull:2013:Optimal},
boundary conditions of $\boldsymbol{z}=(\mu_{1},\mu_{2},\gamma_{1},\gamma_{2},\rho_{12})$
and $\boldsymbol{\lambda}$ are as follows: $\mu_{1}(0)=0$, $\mu_{2}(0)=0$,
$\gamma_{1}(0)=0$, $\gamma_{2}(0)=0$, $\rho_{12}(0)=0$, $\mu_{2}(T)=\mu_{2}^{\mathrm{trg}}$,
$\lambda_{1}(T)=0$, $\lambda_{3}(T)=0$, $\lambda_{4}(T)=1$, and
$\lambda_{5}(T)=0$. We numerically solved this two-point boundary
value problem with a \texttt{bvp4c} function in \emph{Matlab}.

\section{Optimal signal dependence on $\beta_{1}$}

\begin{figure}
\begin{centering}
\includegraphics[height=6cm]{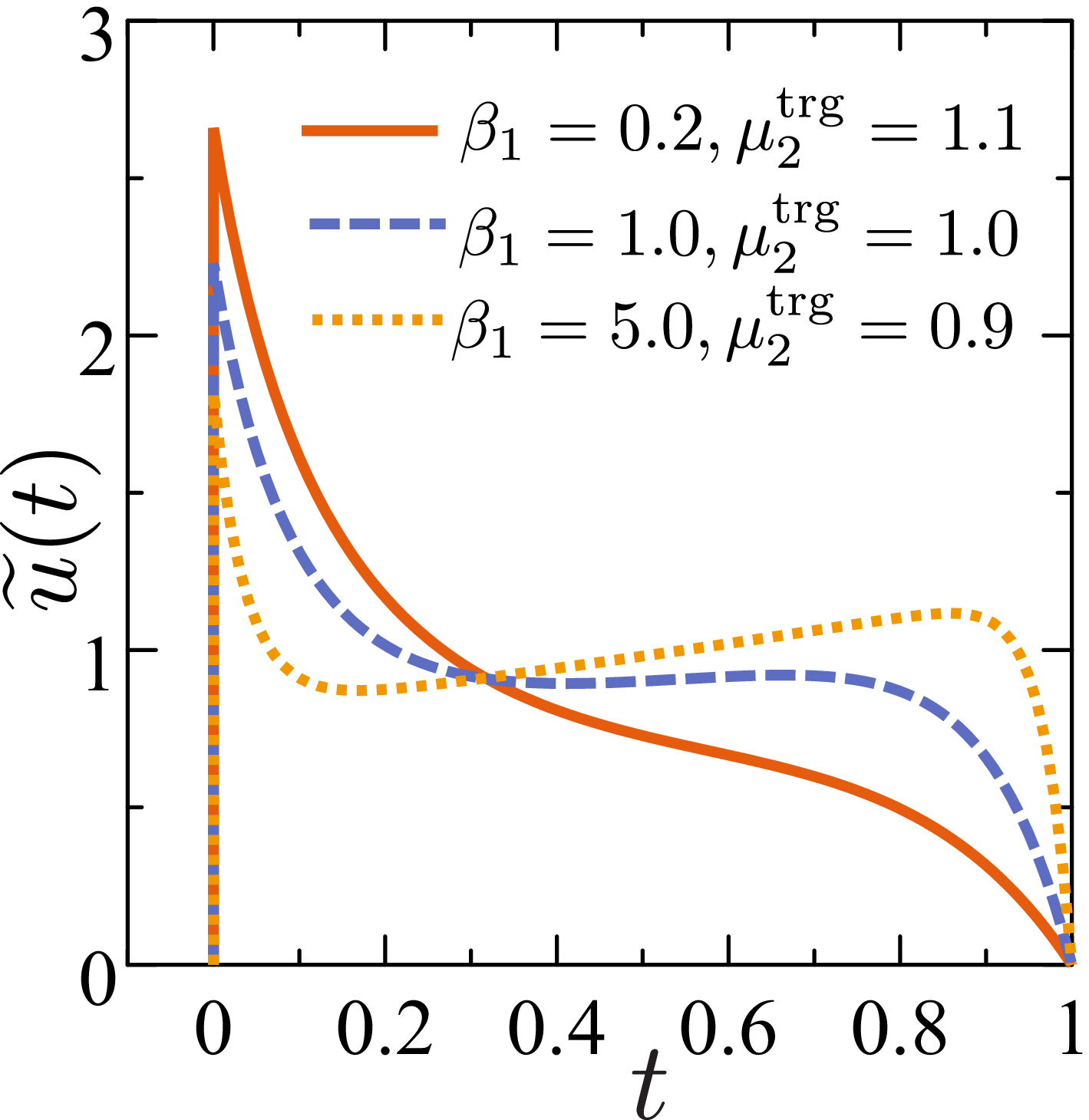}
\par\end{centering}

\protect\caption{Normalized signal $\widetilde{u}(t)$ for three values of $\beta_{1}$:
$\beta_{1}=0.2$ and $\mu_{2}^{\mathrm{trg}}=1.1$ (solid line), $\beta_{1}=1.0$
and $\mu_{2}^{\mathrm{trg}}=1.0$ (dashed line), and $\beta_{1}=5.0$
and $\mu_{2}^{\mathrm{trg}}=0.9$ (dotted line). We set $w=1.0$ and
the unspecified parameters are the same as those in Fig.~\FIGIIIDPlotI(a).
\label{fig:fig_beta1}}
\end{figure}

In the main text, we studied shape of the optimal signal dependence
on $\beta_{2}$ (Section~3.2). We performed the same calculation
for $\beta_{1}$, i.e. $\beta_{1}$ and $\mu_{2}^{\mathrm{trg}}$
are varied while keeping other parameters unchanged (the other parameters
are the same as those in Fig.~\FIGIIIDPlotI(a)). Figure~\ref{fig:fig_beta1}
shows normalized signal $\widetilde{u}(t)=u(t)/\sqrt{\widetilde{\Pi}}$
for three cases: $\beta_{1}=0.2$ and $\mu_{2}^{\mathrm{trg}}=1.1$
(solid line), $\beta_{1}=1.0$ and $\mu_{2}^{\mathrm{trg}}=1.0$ (dashed
line), and $\beta_{1}=5.0$ and $\mu_{2}^{\mathrm{trg}}=0.9$ (dotted
line). We found that all the optimal signals exhibit the overshooting.
This implies that the relaxation time of $X_{1}$ is not responsible
for the overshooting pattern. 

\providecommand{\newblock}{}